\documentclass[prl,twocolumn,10pt,aps,superscriptaddress]{revtex4-1}
\pdfoutput=1

\usepackage{amsmath,amsthm,amssymb,bm,hyperref,graphicx,color}
\usepackage{times}

\newcommand{\nc}{\newcommand}

\newcommand{\zcor}[2]{\bra \sigma_{#1}^z \sigma_{#2}^z \ket}
\nc{\bra}{\langle}
\nc{\ket}{\rangle}
\nc{\omd}[2]{{(#1,#2)}}
\definecolor{red}{rgb}{1,0,0}

\def\beq{\begin{equation}}
\def\eeq{\end{equation}}
\def\bea{\begin{eqnarray}}
\def\eea{\end{eqnarray}}
\def\vspec{x}
\def\vspecy{y}
\def\hspec{\epsilon}
\let\nn=\nonumber
\def\beann{\begin{eqnarray*}}
\def\eeann{\end{eqnarray*}}

\let\a=\alpha

    \let\s=\sigma

\theoremstyle{plain}

\newtheorem*{corollary*}{Corollary}

\newtheorem*{conjecture*}{Conjecture}

\theoremstyle{definition}

\def\2{\frac{1}{2}} \def\4{\frac{1}{4}}

\def\6{\partial}

\def\+{\dagger}

\def\<{\langle} \def\>{\rangle}

\def\CH{{\cal H}}

\def\i{{\rm i}}

\def\jg2{K}

\renewcommand{\appendix}{%
   \renewcommand{\section}{
        \secdef\Appendix\sAppendix}%
   \setcounter{section}{0}%
   \renewcommand{\thesection}{\Alph{section}}%
   \renewcommand{\theequation}{\thesection.\arabic{equation}}%
}

\newcommand{\Appendix}[2][?]{%
     \refstepcounter{section}%
     \setcounter{equation}{0}%
     \addcontentsline{toc}{appendix}%
          {\protect\numberline{\appendixname~\thesection} #1}%
     \vspace{\baselineskip}%
     {\noindent\large\bfseries\appendixname\ \thesection: #2\par}%
     \sectionmark{#1}\vspace{\baselineskip}}

\newcommand{\sAppendix}[1]{%
     {\noindent\large\bfseries\appendixname\:: #1\par}%
     \sectionmark{#1}\vspace{\baselineskip}}

\begin{document}

\title{Computation of static Heisenberg chain correlators: Control over
length and temperature dependence}

\author{Jun Sato}
\affiliation{Ochanomizu University, Otsuka 2-1-1, Bunkyo-ku, Tokyo, 112-8610, Japan}
\author{Britta Aufgebauer,
Herman~Boos,
Frank G\"{o}hmann,
Andreas Kl\"umper,
Minoru Takahashi,
Christian Trippe}
\affiliation{Fachbereich C -- Physik, Bergische Universit\"at Wuppertal,
42097 Wuppertal, Germany}

\begin{abstract}
We communicate results on correlation functions for the spin-1/2 Heisenberg
chain in two particularly important cases: (a) for the infinite chain at
arbitrary finite temperature $T$, and (b) for finite chains of arbitrary
length $L$ in the ground state. In both cases we present explicit formulas
expressing the short-range correlators in a range of up to seven lattice sites
in terms of a single function~$\omega$ encoding the dependence of the
correlators on $T$ ($L$). These formulas allow us to obtain accurate numerical
values for the correlators and derived quantities like the entanglement
entropy. By calculating the low $T$ (large $L$) asymptotics of $\omega$ we
show that the asymptotics of the static correlation functions at any finite
distance are $T^2$ ($1/L^2$) terms. We obtain exact and explicit formulas for
the coefficients of the leading order terms for up to eight lattice sites.
\end{abstract}

\maketitle

Until about ten years ago it was widely believed that it would be
practically impossible to calculate the lattice correlation functions of the
Heisenberg chain explicitly, despite the fact that its Hamiltonian
\begin{equation}\label{eq:ham}
     \CH_N = J \sum_{j = 1}^L 
           \vec\s_{j} \vec\s_{j+1}
\end{equation}
is one of the best known among those integrable lattice systems whose spectrum
can be exactly calculated by means of the Bethe ansatz \cite{Bethe31}. Starting
from about the year 2000, however, our understanding of the model has changed
profoundly.

The recent progress is due to several different but related methods, like the
representation theory of quantum algebras, the algebraic Bethe ansatz, or the advent
of new functional equations, which first of all led to the derivation of multiple
integral representations \cite{JMMN92,*JiMi96,*KMT99b,*GKS04a} for the elements
of the density matrix. In a series of works \cite{BoKo01,*BKS03,*BKS04c,BJMST04a}
it was then shown that the correlation functions of an inhomogeneous version of the
model can be represented by algebraic expressions of a single function of two variables.

An exponential formula for the reduced density matrix was obtained in
\cite{BJMST05b,*BJMST06}. This formula will be of particular importance below,
since it is also valid for finite temperature or finite length as was
conjectured in \cite{BGKS06,*DGHK07}. In \cite{BJMST06b,*BJMST08a} a fermionic
structure on the space of local operators was identified from which a generating
function of all local correlation functions was obtained \cite{JMS08} for a
very general inhomogenous vertex model including the finite temperature and
the finite length Heisenberg chain as special cases. Even in this most general
situation the inhomogeneous correlation functions depend on only two functions
\cite{JMS08,BoGo09}. Apart from these important theoretical findings many
results for the ground state correlation functions in the thermodynamic limit,
but also a number of finite temperature or finite length correlation functions
were obtained explicitly, see e.g.\ \cite{KMST05,*TGK10a} and the references therein. 

The aim of this paper is twofold. First, we present recent exact results for
the temperature and size dependence of two-point correlation functions. Of
particular interest are system parameters $T$ and $L$ in the regime of
conformal field theory (CFT) and their influence on short-ranged correlators
that are --strictly speaking-- outside the domain of CFT. Furthermore, we
present results for the entanglement entropy of sub-segments of an infinitely
long chain in the entire temperature window from 0 to $\infty$. Second, we
intend to explain the necessary computations in the light of a recent
understanding of the subject \cite{Aufgebauer11} based on `discrete functional
equations' which we believe is physically most natural.

All information about the correlation functions of operators acting on $n$
successive sites of the chain are encoded in the reduced density matrix $D_n$.
The matrix elements of $D_n$ are denoted by
$D_{\sigma_1,...,\sigma_n}^{\mu_1,...,\mu_n}$ and correspond to the
expectation values of operators
$|\sigma_1,...,\sigma_n\rangle\langle\mu_1,...,\mu_n|$.

A quantum mechanical system at finite temperature $T$ may be viewed as a
classical statistical system on a cylinder of circumference
$\beta=1/T$. In this way, the density matrix $D_n$ on $n$ successive sites for
temperature $T$ is obtained in a suitable limit of the six-vertex model
density matrix on a rectangular lattice of unbounded width and finite height
$N$. Each row corresponds to an imaginary time slice of height $\tau=\beta/N$,
the continuous time limit is obtained in the Trotter limit $N\to\infty$. For
our purposes it is convenient to have independent heights $\tau_1,...,\tau_N$
(resp.~spectral parameters $\hspec_1$,... $\hspec_N$ placed on the horizontal
lines) under suitable conditions like $\tau_j={\cal O}(1/N)$ and
$\sum_j\tau_j=\beta$.

We introduce independent spectral parameters $\vspec_1,...,\vspec_n$ on the
vertical lines corresponding to the $n$ sites picked for the definition of
$D_n$.  The density matrix of the generalized problem now depends on the
${\vspec_j}$'s and is denoted by $D_n(\vspec_1,...,\vspec_n)$ with matrix
elements
$D_{\sigma_1,...,\sigma_n}^{\mu_1,...,\mu_n}(\vspec_1,...,\vspec_n)$. The full
functional dependence will be solved, the subsequent specialization of the
arguments yields the physically interesting data. The reason behind the
success of this solution strategy is the analyticity of the object as a
function of the spectral parameters $\vspec$. This is a consequence of the
integrabilty of the system: the eigenstates of the column-to-column transfer
matrices $T(\vspec)$ do not depend on $\vspec$ and hence the eigenvalues and
other objects as functions of $\vspec$ do not show the otherwise unavoidable
root singularities.

As an example of this solution strategy we like to note results for the
leading eigenvalue $\Lambda(\vspec)$ of the transfer matrix
$T(\vspec)$ which will be of use below
\begin{equation}
\log\Lambda(\vspec)=e_0(\vspec)
+\frac 1{2\pi}\int_{{\cal C}} \frac{\log(1+a(\vspecy))}{(\vspec-\vspecy-\i)
(\vspec-\vspecy)}d\vspecy.\label{eigenvalue}
\end{equation}
Here, ${\cal C}$ is a narrow closed contour around the real axis, $e_0$ is
some elementary function containing the $\hspec_j$-parameters.
The auxiliary function $a(\vspec)$ satisfies the non-linear integral equation
\begin{equation}
\log a(\vspec)= a_0(\vspec)-\frac 1{\pi}\int_{{\cal C}} 
\frac{\log(1+a(\vspecy))}{1+(\vspec-\vspecy)^2}d\vspecy,\label{NLIEa}
\end{equation}
where $a_0(\vspec)$ is an elementary function of $\vspec$ and the $\hspec_j$,
which for finite temperature, in the Trotter limit $N\to\infty$, takes the
form $a_0(\vspec)=2J\beta/(\vspec(\vspec+\i))$. For the ground-state and
finite size it is simply $a_0(\vspec)= L \log\frac{\vspec-\i/2}{\vspec+\i/2}$.

The program of calculating the functional dependence of
$D_n(\vspec_1,...,\vspec_n)$ is well understood in the case of zero
temperature. For $T=0$ (i.e.~$\beta=\infty$) the object
$D_n(\vspec_1,...,\vspec_n)$ satisfies non-trivial linear functional
equations in all arguments. The most important equation, the `rqKZ'-equation
reads for instance
\begin{equation}
\mspace{-6.mu}
D_n(\vspec_1,...,\vspec_{n-1},\vspec_n-1)=
A(\vspec_1,...,\vspec_n)D_n(\vspec_1,...,\vspec_n) \label{rqKZ}
\end{equation}
where $A$ is a linear operator \cite{BJMST04a} acting in the space of
density matrices and the $\vspec_j$ may take any value from the complex
plane. The action of $A$ on $D_n$ is of the following type
\begin{equation}
(A D)_{\vec\sigma}^{\vec\mu}
  =\sum_{\vec\sigma',\vec\mu'}A_{\vec\sigma,\vec\mu'}^{\vec\mu,\vec\sigma'}
D_{\vec\sigma'}^{\vec\mu'}
\end{equation}

The functional equations can be solved for the analytical function $D$. The
uniqueness of the solution for $T=0$ is guaranteed by the asymptotic behaviour for
$\vspec_j\to\infty$. The finding of \cite{BJMST04a} is striking, a two-point
nearest neighbour function $\omega(\vspec_1,\vspec_2):=$ \hbox{$6$ tr
  $D_2(\vspec_1,\vspec_2)S_1^zS_2^z$} and a set of `structure constants'
$f_{n,I,J}$ determine the density matrix $D_n$ for arbitrary $n$
\begin{equation}
\mspace{-9.mu}
D_n=\sum_{m=0}^{[n/2]}\sum_{I,J}\left(\prod_{p=1}^{m}\omega(\vspec_{I_p}, 
\vspec_{J_p})\right)
f_{n,I,J}(\vspec_1,\vspec_2,...,\vspec_n)
\label{factorizedExp}
\end{equation}
The summation labels $I$ and $J$ are $m$-tuples of integers such that $I\cap
J=\emptyset$ and $I_1<\dots<I_m,\,1\le I_p<J_p\le n$.
The structure coefficients
$f_{n,I,J}(\vspec_1,\vspec_2,...,\vspec_n)$ are
matrices with elementary rational functions of the arguments
$\vspec_1,\vspec_2,...,\vspec_n$ as entries. For details see \cite{BJMST04a}.

At zero temperature the computation of the correlation functions employed two
important facts: (i) the functional equation (\ref{rqKZ}) is satisfied for
continuous arguments, and (ii) $D_n(\vspec_1,...,\vspec_n)$ depends on the
arguments only via the differences $\vspec_j-\vspec_i$.  This is fundamentally
different for finite temperature, i.e.~finite Trotter number $N$. The main
problem is that (\ref{rqKZ}) no longer holds: on
the r.h.s. of (\ref{rqKZ}) untreatable `correction terms' appear, a serious
obstacle so far for treating $T>0$.

At this point, the computation of the density matrix for finite Trotter number
$N$ on the basis of functional equations looks unfeasable.  However, for the
above six-vertex model with $N$ many rows carrying spectral parameters
$\hspec_1,...,\hspec_N$ we find \cite{Aufgebauer11} a discrete version of
functional equations. In detail:

\noindent
(I) Equation (\ref{rqKZ}) `rqKZ' holds literally (!) if we restrict
$\vspec_n$ to the set $\{\hspec_1,...,\hspec_N\}$.\\
(II) A reduction takes place in the limit of large spectral parameter $\vspec_n$,
i.e.~$D_n(\vspec_1,...,\vspec_n)\to
D_{n-1}(\vspec_1,...,\vspec_{n-1})\otimes D_1$ where $D_1$ is a single site
density matrix of a paramagnetic spin. 
For zero magnetic field this is $D_1=\frac12$id.\\
(III) Analyticity properties:
\begin{equation}
D_{\sigma_1,...,\sigma_n}^{\mu_1,...,\mu_n}(\vspec_1,...,\vspec_n)
=\frac{P_{\sigma_1,...,\sigma_n}^{\mu_1,...,\mu_n}(\vspec_1,...,\vspec_n)}
{\Lambda(\vspec_1)\cdot...\cdot\Lambda(\vspec_n)}
\end{equation}
where $P_{\sigma_1,...,\sigma_n}^{\mu_1,...,\mu_n}(\vspec_1,...,\vspec_n)$
is an $n$-variate polynomial of degree $N$ in the variables
$\vspec_1,...,\vspec_n$ and $\Lambda(\vspec)$ is the largest eigenvalue of
the transfer matrix $T(\vspec)$ obtained from (\ref{eigenvalue}).
 
The computational problem consists in finding the polynomials
$P_{\sigma_1,...,\sigma_n}^{\mu_1,...,\mu_n}(\vspec_1,...,\vspec_n)$. We
have proved \cite{Aufgebauer11} that the above equations have a unique solution. Hence
any expression (or ansatz) for $D_n(\vspec_1,...,\vspec_n)$ exhibiting the
above listed properties realizes {\em the} solution.

We find that (I) is satisfied by (\ref{factorizedExp}) like in \cite{BJMST04a}
with the same structure coefficients $f_{n,I,J}$ provided that
$\omega(\vspec_1,\vspec_2)$ is a symmetric function satisfying the `discrete
functional equation' (see \cite{BoGo09} for the most general case)
\begin{equation}
\frac{\omega(\vspec_1,\vspec_2)}{\vspec^2-1}+
\frac{\omega(\vspec_1,\vspec_2-1)}{\vspec(\vspec+2)}=\frac3{2(\vspec^2-1)\vspec(\vspec+2)}
\end{equation}
where $\vspec:=\vspec_1-\vspec_2$, $\vspec_1$ is arbitrary and $\vspec_2\in\{\hspec_1,...,\hspec_N\}$. The asymptotics (II) is
satisfied if $\omega(\vspec_1,\vspec_2)\to 0$ for $\vspec_1\to\infty$ or
$\vspec_2\to\infty$.  Condition (III) is satisfied if $\omega$ is a polynomial
(of degree $N-1$) divided by $\Lambda(\vspec_1)\Lambda(\vspec_2)$. These
conditions for $\omega$ characterize a unique function. It is
relatively straightforward to see that the conditions for $\omega$ are
satisfied by the following expressions
\begin{align}
\omega(\vspec_1,\vspec_2) & :=\frac12+
\frac12((\vspec_1-\vspec_2)^2-1)\psi(\i\vspec_1,\i\vspec_2) \notag \\
\psi(\vspec_1,\vspec_2) & :=
\frac 1{\pi}\int_{{\cal C}} \frac{d\vspecy}{1+a(\vspecy)}
\frac{G(\vspecy,\vspec_1)}{(\vspecy-\vspec_2)(\vspecy-\vspec_2-\i)}
\label{omega}
\end{align}
where the function $G$ satisfies the linear integral equation
\begin{multline}
G(\vspec,\vspec_1) = \\ \mspace{-9.0mu} -\frac1{(\vspec-\vspec_1)(\vspec-\vspec_1-\i)}
+\int_{{\cal C}} \frac{d\vspecy/\pi}{1+a(\vspecy)}
\frac{G(\vspecy,\vspec_1)}{1+(\vspec-\vspecy)^2}
\label{linintG}
\end{multline}
and the function $a(\vspec)$ was introduced in (\ref{NLIEa}).

Note that the expression in (\ref{factorizedExp}) contains rational functions
as prefactors. The poles of these coefficients are cancelled by zeros
appearing due to the pairwise cancellation of terms in the sum. Consequently,
the only poles on the r.h.s. of (\ref{factorizedExp}) are those occurring in
the product ${\Lambda(\vspec_1)\cdot...\cdot\Lambda(\vspec_n)}$. Hence, also
(III) is satisfied.

{\em Finite temperature.}
The coefficients in (\ref{factorizedExp}) neither depend on $N$ nor on the
parameters $\hspec_j$. These only enter in $\omega$ via the dependence of $a$
and $G$ on them. Therefore, taking the Trotter limit $N\to\infty$ is easy. 
\begin{figure}
  \begin{center}
    \includegraphics[width=\linewidth,angle=0]{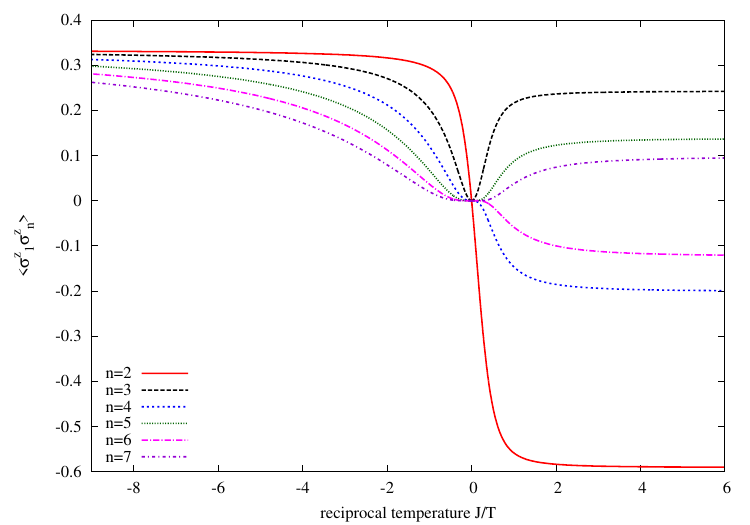}
    \caption{Depiction of two-point correlators $\bra \sigma_{1}^z \sigma_{n}^z \ket$ for
different point separations $n$ in dependence on $J/T$ for antiferromagnetic
(ferromagnetic) exchange to the right (left) of 0.
}
    \label{fig:corr_af_f}
  \end{center}
\end{figure}

The evaluation of the correlation function consists in numerically solving (\ref{NLIEa})
for $a(x)$. Then (\ref{linintG}) is
solved for $G$ and eventually $\omega$ is calculated from (\ref{omega}).
The definition of the coefficients in (\ref{factorizedExp}) may be
found in \cite{BJMST04a}, but an efficient computation is due to
\cite{SST05,*Takahashi11}. Here we use their results for density matrices $D_n$
with $n$ up to 7 sites, and combine this with the finite temperature results for the
function $\omega$. The formulas for $D_n(\vspec_1,...,\vspec_n)$ allow for
taking the homogeneous limit $\vspec_1=...=\vspec_n=0$ and yield simple
algebraic expressions involving $\omega$ and its derivatives.

Just as an illustration of the explicit expressions we like to show formulas
for the 2-point correlation functions obtained from $D_n$ for $n=2, 3, 4$
\begin{align}
\zcor{1}{2}&=\frac{2}{3} \omd{0}{0},\ 
\zcor{1}{3}=\frac{2}{3} \omd{0}{0} + \frac{2}{3} \omd{1}{1} - \frac{1}{3} \omd{2}{0}
\notag \displaybreak[0] \\
\zcor{1}{4}&=\omd{0}{0}\left[\frac{2}{3}+ \frac{4}{3} \omd{1}{1}
+ \frac{2}{9} \omd{2}{2} - \frac{4}{27}\omd{3}{1}\right]\notag \\ & \mspace{-18.mu}
-\omd{1}{0}\left[\frac{4}{3} \omd{1}{0}+\frac{4}{9}\omd{2}{1}
-  \frac{4}{27} \omd{3}{0}\right]- \frac{1}{9} \omd{3}{1}\nn\\ & \mspace{-18.mu}
+\left[4\omd{1}{1}-{2}\omd{2}{0}\right]\left[\frac{1}{3}
+\frac{1}{9} \omd{2}{0}\right] + \frac{1}{6} \omd{2}{2}
\label{CorrOmega}
\end{align}
where
$(j,k):=\partial^j_1\partial^k_2\omega(\vspec_1,\vspec_2)|_{\vspec_1=\vspec_2=0}$.
We derived similar formulas for the cases $n=5,...,8$ but these are by far too
long to be shown here.
The correlators $\zcor{1}{n}$ are analytic functions of $J/T$ along the entire
real axis with zero of $(n-1)$th order at $J/T=0$. 
In Fig.~\ref{fig:corr_af_f} we
show the results for the antiferromagnetic chain: data at negative values of $J/T$
correspond to results for the ferromagnetic chain at $|J/T|$.
In the
ferromagnetic case the correlations are strictly positive, in the antiferromagnetic
case, the correlations are negative (positive) for even (odd) $n$.

In \cite{CGK10} the low-temperature behaviour of $\omega(\vspec_1,\vspec_2)$
was calculated with the result
\begin{equation}
\omega=\omega_0+\frac{T^2}{24
  J^2}\left(1-(\vspec_1-\vspec_2)^2\right)\cos(\pi(\vspec_1+\vspec_2)),
\label{omegalowT}
\end{equation}
where $\omega_0$ denotes the $T=0$ limit of $\omega$ in the thermodynamic limit.
From this expansion and (\ref{CorrOmega}) we obtain explicit results for the
correlations like
\begin{align}
\mspace{-9.mu}
\zcor{1}{2}&\simeq\frac 13-\frac 43\ln 2+\frac 1{36}(T/J)^2 \notag \\
\mspace{-9.mu}
\zcor{1}{3}&\simeq\frac 13-\frac {16}3\ln 2+3\zeta(3)+\left(\frac19-\frac{\pi^2}{72}\right)(T/J)^2
\end{align}
Similar but more lengthy expressions up to $\zcor{1}{8}$ have been
derived. These results can be written in the following form
\begin{equation}
\zcor{1}{1+r}\simeq\zcor{1}{1+r}_{0}(1-\gamma_r(T/J)^2)\label{lowTexpansionCorr}
\end{equation}
where explicit numbers for $\gamma_r$ are given in table \ref{my_table}.
\setlength{\tabcolsep}{0pt}
\begin{table}
\begin{ruledtabular}
\begin{tabular}{ccccccc}
1 & 2 & 3 & 4 & 5 & 6 & 7  \\
0.0470 & 0.1070 & 0.3268 & 0.5014 &
0.9013 & 1.1957 & 1.7761\\
1.1283 & 0.6419 & 0.8714 &
0.7521 & 0.8652 & 0.7971 & 0.8699\\
\end{tabular}
\end{ruledtabular}
\caption{The low-temperature expansion coefficients $\gamma_r$ of the
  correlations $\zcor{1}{1+r}$ for $r=1,...,7$ (2nd row) and the ratio of $\gamma_r$
  with the CFT prediction $\gamma_r^{CFT}=r^2/24$ (3rd row).}
\label{my_table}
\end{table}

For a (primary) field with scaling dimension $x$ the two-point correlator at
distance $r$ is given by
\begin{equation}
C_r(T)= C \biggl(\frac{\pi T/{v}}{\sinh\pi r T/{v}}\biggr)^{2x}
    \simeq\frac{C}{r^{2x}} \left(1-\frac {x}{3}(\pi r T/v)^2\right)
\end{equation}
where the sound velocity of the elementary excitations is $v=2\pi J$. For
the spin-spin correlations we have to set $x=1/2$. Hence, the CFT prediction
of the coefficient in (\ref{lowTexpansionCorr}) is $\gamma_r^{CFT}=r^2/24$. In
table \ref{my_table} results for the ratio $\gamma_r/\gamma_r^{CFT}$ are
given. The values are of the order 1, but deviate significantly from 1 which
we attribute to the simple form of the CFT prediction (\ref{lowTexpansionCorr})
which is strictly valid only for conformally invariant models without
marginally irrelevant perturbations which exist in the isotropic
Heisenberg chain. Also, it is likely that the sequence
$\gamma_r/\gamma_r^{CFT}$ has two different accumulation points for even and
odd $r$, respectively.

\begin{figure}
  \begin{center}
    \includegraphics[width=\linewidth,angle=0]{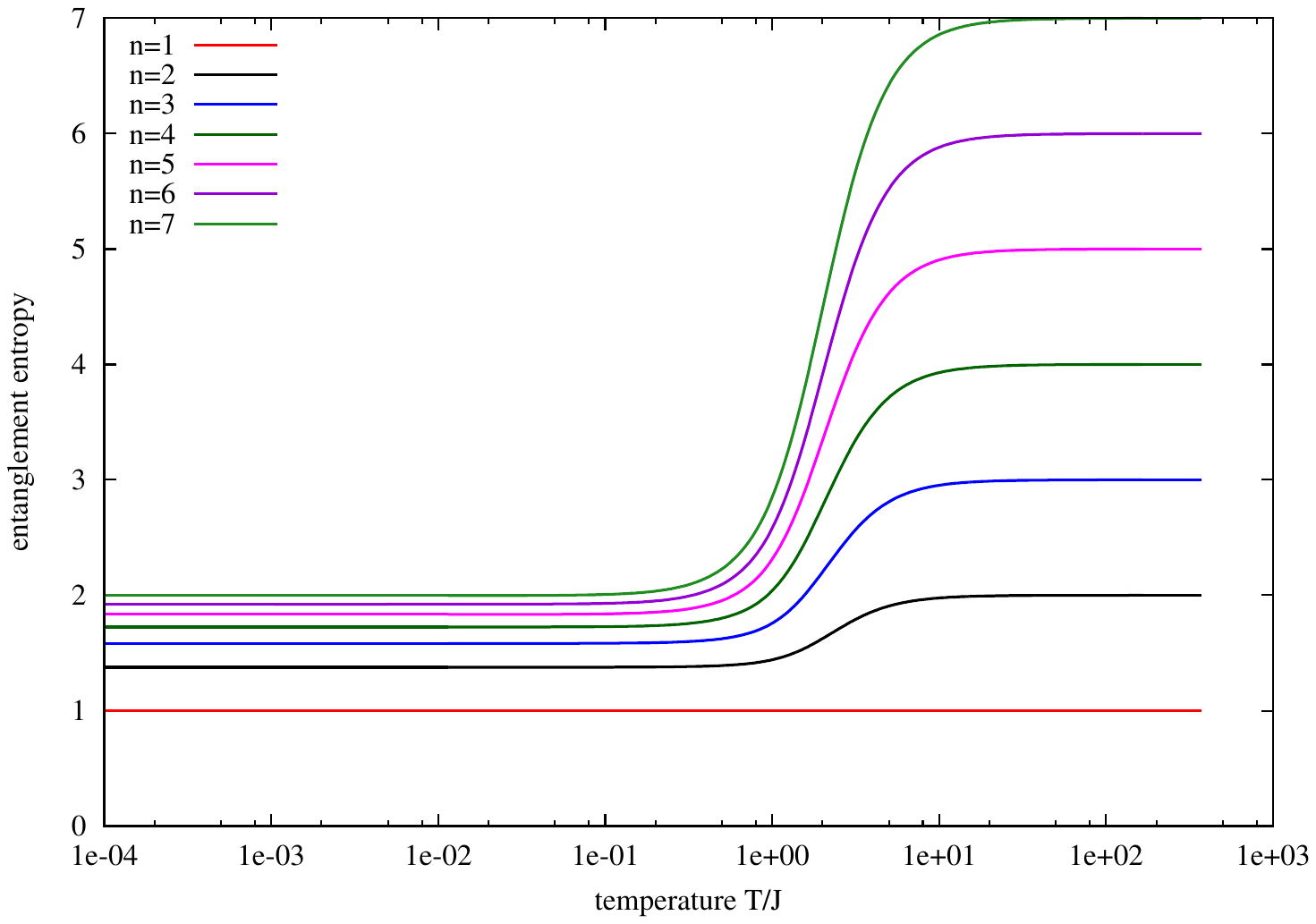}
    \caption{Depiction of the entanglement entropies of $D_n$ as functions of
      temperature.}
    \label{fig:entropies}
  \end{center}
\end{figure}

In Fig.~\ref{fig:entropies} we show the entanglement entropies $S_n(T)$ for
blocks of size $n$ for $n=1,...,7$. Note that we used the logarithm with base
2 (the total number of local states) in the definition of the entropy. Hence
the high temperature asymptote of $S_n$ is identical to the length of the
block $n$. The low-temperature limit scales with $1/3 \log_2(n)$.%
\begin{figure}
  \begin{center}
    \includegraphics[width=\linewidth,angle=0]{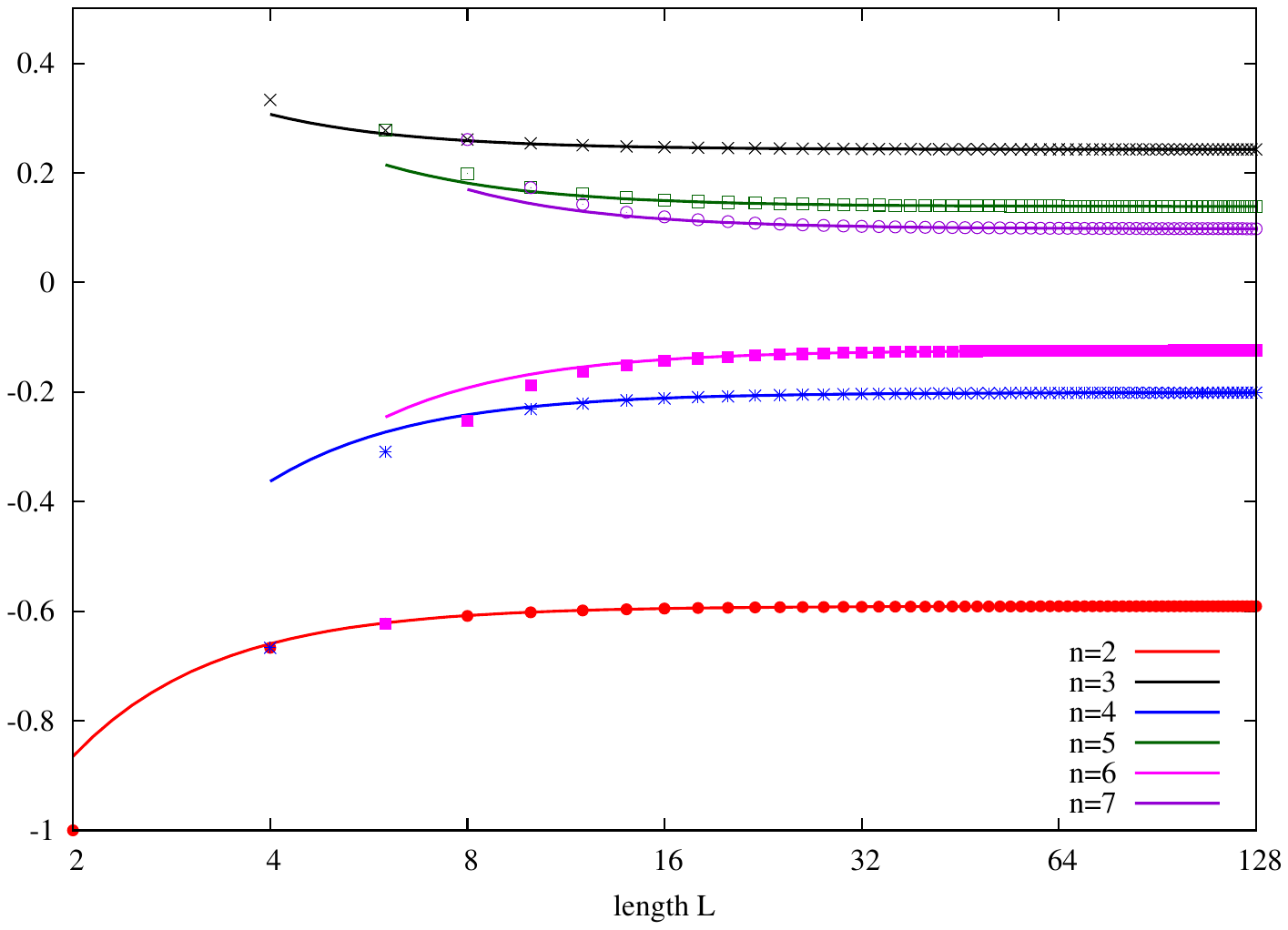}
    \caption{Antiferromagnetic Heisenberg chain at $T=0$: Depiction of 2-point
      correlators $\zcor{1}{n}$ in dependence on chain length
      $L$ for different point separations $n$. Solid lines are finite size fits.}
    \label{fig:Finitesize}
  \end{center}
\end{figure}%

{\em Finite Size.}
Similar to (\ref{omegalowT}) we find in the large $L$ limit
\begin{equation}
\omega=\omega_0+\frac{\pi^2}{6L^2}\left(1-(\vspec_1-\vspec_2)^2\right)\cos(\pi(\vspec_1+\vspec_2)).
\label{omegalargeL}
\end{equation}
For finite size, the physical density matrix is $D_n=D_n(1/2,...,1/2)$. Hence,
all the above formulas for low $T$ turn into their counterparts for large $L$
if we replace $(2J/T)^2$ by $-L^2/\pi^2$ leading to
\begin{equation}
\zcor{1}{1+r}\simeq\zcor{1}{1+r}_{0}(1+4\gamma_r\pi^2/L^2).\label{largeLCorr}
\end{equation}
In contrast to finite temperature, finite size increases the correlations.
In Fig.~\ref{fig:Finitesize} we show data for chains of length $L$ up to
$128$. Note that correlators $\zcor{1}{n}$, $\zcor{1}{m}$ for the same $L$ coincide
if $n+m=L+2$.

In conclusion we have derived exact results for correlation functions of the
Heisenberg spin chain for finite temperature $T$ (finite system size $L$). In
the conformal regime of low $T$ (large $L$), the corrections to the
ground-state results in the thermodynamic limit are additive $T^2$ ($L^{-2}$)
terms. The exponents are universal and agree with CFT predictions. The
coefficients for two-point correlators at strictly finite lattice separation
are non-universal, but correspond precisely to those quantities that are of
central interest in many, especially numerical approaches like the
diagonalization of Hamiltonians or Monte-Carlo simulations. We also managed to
derive multi-spin correlations and as an example we showed exact data for the
entanglement entropy for arbitrary temperatures with smooth transitions from
$T=0$ to $T=\infty$ with rather different dependence on the length of the
chain segment.  The central expression for all (static) correlation functions
shows a remarkable structure, it is a sum of products of nearest-neighbour
correlators.  We also explained how the central expression for the
correlations can be derived from a set of `discrete' functional equations for
the density matrix.  We are convinced that this method is very powerful and
applicable to other seminal models with higher spins or different symmetry
groups as well.

{\it Acknowledgments.} M.T. acknowledges the {\em Mercator professorship}
financed by DFG.



\bibliographystyle{apsrev4-1}
\bibliography{hub}

\end{document}